\global\def\draftcontrol{0}

   \def\versionno{ vgr}

\catcode`\@=11

\expandafter\ifx\csname draftcontrol\endcsname\relax\global\def\draftcontrol{0}
\fi

{\count255=\time\divide\count255 by 60
\xdef\hourmin{\number\count255}
\multiply\count255 by-60\advance\count255 by\time
\xdef\hourmin{\hourmin:\ifnum\count255<10 0\fi\the\count255}}
\def\draftdate{\number\month/\number\day/\number\year\ \ \ \hourmin }

\newcommand\makepapertitle{\par
  \begingroup
    \renewcommand\thefootnote{\@fnsymbol\c@footnote}%
    \def\@makefnmark{\rlap{\@textsuperscript{\normalfont\@thefnmark}}}%
    \long\def\@makefntext##1{\parindent 1em\noindent
            \hb@xt@1.8em{%
                \hss\@textsuperscript{\normalfont\@thefnmark}}##1}%
     \newpage
     \global\@topnum\z@   
     \@makepapertitle
     \thispagestyle{empty}\@thanks
  \endgroup
  \setcounter{footnote}{0}%
  \global\let\thanks\relax
  \global\let\makepapertitle\relax
  \global\let\@makepapertitle\relax
  \global\let\@thanks\@empty
  \global\let\@author\@empty
  \global\let\@date\@empty
  \global\let\@title\@empty
  \global\let\title\relax
  \global\let\author\relax
  \global\let\date\relax
  \global\let\and\relax
  \def\version{\let\version\@version\@gobble}
}
\def\@makepapertitle{%
  \newpage
   \ifnum\draftcontrol=1 {}
   \version\versionno
   \vskip 3em%
   \else
   \hfill\hbox to 3cm {\parbox{4cm}{\@pubnum}\hss}%
   \vskip 3em%
   \fi
   \begin{center}%
   \let \footnote \thanks
     {\LARGE {\@title}}%
     \vskip 1.5em%
     {\normalsize
       \lineskip .5em%
       \begin{tabular}[t]{c}%
         \@author
       \end{tabular}\par}%
     \vskip 1.5em%
     {\@bstract}%
     \end{center}%
     \vskip 1.5em
     \@date%
   \par
}

\gdef\@pubnum{}
\def\pubnum#1{%
  \gdef\@pubnum{#1}}

\gdef\@bstract{}
\def\Abstract#1{%
  \gdef\@bstract{%
   \parbox{\textwidth-0pc}{%
   \centerline{\bf Abstract}\penalty1000%
\kern.2cm%
\noindent
\renewcommand\baselinestretch{1.0}%
{#1}}}
}

\def\ps@paper{\let\@mkboth\@gobbletwo%
     \ifnum\draftcontrol=1
    \def\@oddfoot{\hbox to \textwidth{\tiny \versionno \hfil\tiny\draftdate}%
    \hskip -\textwidth \hbox to \textwidth{\hfil\rm\thepage\hfil}}%
     \else\def\@oddfoot{\hbox to \textwidth{\hfil\rm\thepage\hfil}}
     \fi
     \let\@evenfoot\@oddfoot
}

\def\body{\clearpage
          \pagestyle{paper}
    }

\def\@version#1{\ifnum\draftcontrol=1
\typeout{}\typeout{#1}\typeout{}
\vskip3mm\centerline{\hbox{\fbox{\normalsize{\tt DRAFT -- #1 -- }
                   {\draftdate}}}}\vskip3mm
\fi}
\let\version\@version
\long\def\eqlabel#1{\ifnum\draftcontrol=1
                    \tag@false  
                    \tag*{(\theequation) \hbox to -0.2cm{\hspace{0cm}\small{#1}\hss}}
                    \refstepcounter{equation}
                    \edef\@currentlabel{\theequation}
                    \ltx@label{#1}          
                    \else
                    \label{#1}
                    \fi
                    }
\let\st@bibitem\@bibitem
\let\st@lbibitem\@lbibitem
\ifnum\draftcontrol=1
  \def\@bibitem#1{%
    \st@bibitem{#1}\a@@label{#1}\ignorespaces}
  \def\@lbibitem[#1]#2{%
    \st@lbibitem[#1]{#2}\a@@label{#2}\ignorespaces}
  \def\a@@label#1{%
    \gdef\a@lab{\smash{\normalfont\small#1}}
    \ifvmode
      \if@inlabel
        \global\setbox\@labels\hbox{%
          \llap{\a@lab\let\a@lab\relax
                \kern\@totalleftmargin\kern\marginparsep}%
          \box\@labels}%
      \fi
    \fi}
\fi

\documentclass[12pt,letterpaper]{article}

\usepackage{amsmath,amssymb,array,calc,epsfig,rotating,bm}
\usepackage[sort]{cite}
\usepackage{graphicx}
\usepackage{psfrag,verbatim}

\ifnum\draftcontrol=1
\fi

\renewcommand\baselinestretch{1.25}
\renewcommand\section{\@startsection {section}{1}{\z@}%
                                   {-3.5ex \@plus -1ex \@minus -.2ex}%
                                   {2.3ex \@plus.2ex}%
                                   {\normalfont\large\bfseries}}
\renewcommand\subsection{\@startsection{subsection}{2}{\z@}%
                                   {-3.25ex\@plus -1ex \@minus -.2ex}%
                                   {1.5ex \@plus .2ex}%
                                   {\normalfont\normalsize\bfseries}}
\renewcommand\subsubsection{\@startsection{subsubsection}{3}{\z@}%
                                   {-3.25ex\@plus -1ex \@minus -.2ex}%
                                   {1.5ex \@plus .2ex}%
                                   {\normalfont\normalsize\it}}
\renewcommand\paragraph{\@startsection{paragraph}{4}{\z@}%
                                   {-3.25ex\@plus -1ex \@minus -.2ex}%
                                   {1.5ex \@plus .2ex}%
                                   {\normalfont\normalsize\bf}}

\numberwithin{equation}{section}

\def\revise#1       {\raisebox{-0em}{\rule{3pt}{1em}}%
                     \marginpar{\raisebox{.5em}{\vrule width3pt\
                     \vrule width0pt height 0pt depth0.5em
                     \hbox to 0cm{\hspace{0cm}{%
                     \parbox[t]{4em}{\raggedright\footnotesize{#1}}}\hss}}}}

\newcommand\nxt[1]  {\\\fnxt#1}
\newcommand{\ie}{{\it i.e.,}\ }

\def\caln         {{\cal N}}

\def\calr         {{\cal R}}

\def\sqr#1#2{{\vcenter{\vbox{\hrule height.#2pt
 \hbox{\vrule width.#2pt height#1pt \kern#1pt
 \vrule width.#2pt}\hrule height.#2pt}}}}

\def\dd{\delta}

\def\aa1{\phi}
\def\cc1{\psi}

\catcode`\@=12

\begin{document}


\title{\bf Verlinde Gravity and AdS/CFT}

\date{February 27, 2017}

\author{
Alex Buchel\\[0.4cm]
\it Department of Applied Mathematics\\
\it Department of Physics and Astronomy\\ 
\it University of Western Ontario\\
\it London, Ontario N6A 5B7, Canada\\
\it Perimeter Institute for Theoretical Physics\\
\it Waterloo, Ontario N2J 2W9, Canada
}

\Abstract{Verlinde argued \cite{Verlinde:2016toy} that baryonic matter in de
Sitter Universe with a Hubble constant $H$ produces a bulk
contribution to the vacuum entanglement entropy $\delta S_{ent}$.
Using the de Sitter temperature $T_{dS}=\frac{H}{2\pi}$, he
interpreted $\delta M$ in the first law $T_{dS}\ \delta
S_{ent}=-\delta M$ with the Dark Matter energy.  We use standard tools
of the gauge/gravity correspondence to present some evidence for the
Verlinde's proposal.
}

\makepapertitle

\body

\version\versionno

Lessons from the gauge/gravity correspondence \cite{m1,Aharony:1999ti} suggest
that gravity is holographic. In \cite{Verlinde:2016toy} Erik Verlinde 
argued that there is also a bulk contribution to the entropy in quantum gravity $+$ matter,
originated from the entanglement of the baryonic matter with the gravitational 
de Sitter vacuum. The entanglement occurs in the presence of the de Sitter thermal bath 
of a temperature $T_{dS}$, and thus, leads to a new dark energy component $\delta M$ 
according to 
\begin{equation}
T_{dS}\ \delta S_{ent} = -\delta M\,.
\eqlabel{dm}
\end{equation}  
He interpreted $\delta M$ with the energy component due to Dark Mater (DM) in $\Lambda$CDM 
cosmology. Next, he proposed elasticity/gravity correspondence to derive a specific 
prediction between a spherical distribution of the baryonic matter, and the
corresponding entanglement-induced distribution of DM. While the proposal provides 
a derivation of the baryonic Tully-Fisher scaling relation for localized 
baryonic matter \cite{McGaugh:2000sr},  it faces challenges explaining observed 
finite-size galaxy rotation curves \cite{Lelli:2017sul,Hees:2017uyk}. 
This is probably not surprising given the heuristic aspects of the Verlinde's 
elasticity/gravity correspondence.  

In this note we use the standard tools of the gauge theory/gravity correspondence 
\cite{m1,Aharony:1999ti} to motivate: 
\nxt where does $\dd S_{ent}$ come from? 
\nxt what is the corresponding temperature $T=T_{dS}$? 
\nxt why does only the baryonic matter (and not the radiation) lead to $\dd S_{ent}$? \\
Much like in \cite{Verlinde:2016toy}, we will not answer why the first law \eqref{dm}
is valid, and why $\dd M$ has to be identified with the DM.  
In fact, (almost) all the relevant computations to address above questions 
have already been presented in \cite{Buchel:2017pto}. We present here the interpretation,
and refer for the technical details and the full list of references to the latter paper.  

\begin{figure}[t]
\begin{center}
\psfrag{y}{{$\frac{m^2}{H^2}$}}
\psfrag{z}{{$\frac{16\pi}{N^2}\ \calr$}}
  \includegraphics[width=4in]{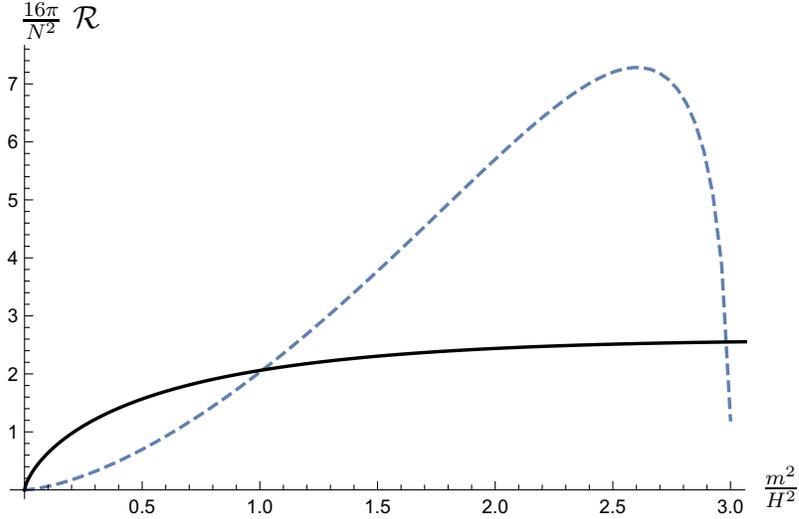}
\end{center}
 \caption{Comoving entropy production rate $\calr$ (see eq.~\eqref{dsdtlate}) of $\caln=2^*$ gauge theory in the $dS_4$ vacuum
 for select values of $\{m_b,m_f\}$: $\{m_b=m_f\equiv m\}$ (solid black curve) and $\{m_f=0,m_b\equiv m\}$ 
(dashed blue curve).}\label{figure1}
\end{figure}

Following \cite{Buchel:2017pto}, a de Sitter vacuum of a strongly coupled $\caln=2^*$ 
gauge theory has a holographic dual gravitational metric 
\begin{equation}
ds_5^2=2 dt\ (dr -A_v\ dt) +\sigma_v^2\ a(t)^2\ d\boldsymbol{x}^2\,, \qquad a(t)=e^{Ht}\,,
\eqlabel{EFmetric}
\end{equation}
where $A_v$ and $\sigma_v$ are functions of a holographic radial coordinate $r$,
and $H$ is a Hubble constant of the de Sitter background.  
The dual geometry \eqref{EFmetric} has an apparent horizon (AH) at $r=r_{AH}$, where
\begin{equation}
H \sigma_v+A_v \sigma'_v\bigg|_{r=r_{AH}}=0 \,.
\eqlabel{AHlate}
\end{equation} 
AH carries Bekenstein-Hawking entropy density
\begin{equation}
s_{BH}=\frac{N^2\Sigma^3}{16\pi}\bigg|_{r=r_{AH}}\,,\qquad  \Sigma\equiv a(t)\ \sigma_v\,,
\eqlabel{sbh}
\end{equation}
which is identified with the comoving entropy density of the boundary 
gauge theory state, \ie $s_{BH}=a^3 s$.
It was argued in \cite{Buchel:2017pto} that this vacuum can not be adiabatic, 
because the comoving entropy production rate is non-zero:
\begin{equation}
\lim_{t\to \infty}\ \frac{1}{H^3 a^3}\frac{d}{dt} (a^3 s)\equiv 3 H\times \calr\,,
\eqlabel{dsdtlate}
\end{equation}
where the rate $\calr$ depends on the mass scales of $\caln=2^*$ gauge theory. It was computed numerically 
for two particular holographic RG flows: $\{m_f=0,m_b\ne 0\}$ and for $\{m_f=m_b\}$, see Fig.~\ref{figure1}.

We would like to interpret the massive $\caln=2^*$ gauge theory as a 'baryonic matter' in Verlinde's theory,  
distributed over {\it all} the de Sitter background space-time. 
If so, the non-adiabaticity of the de Sitter $\caln=2^*$ vacuum \eqref{dsdtlate} can be explained 
by postulating that the entropy density $s$ of the $\caln=2^*$ de Sitter vacuum state (at late times $t\to \infty$)
is precisely Verlinde's entanglement entropy density $s_{ent}$:
\begin{equation}
s_{ent}\equiv \lim_{t\to \infty } s = {H^3}\ \calr\,. 
\eqlabel{senddef}
\end{equation} 
It is natural to call then the dual AH \eqref{AHlate} the {\it entanglement horizon}. 
Notice that since $\calr$ vanishes in the conformal limit, \ie when $\frac{m}{H}\to 0$, 
as in the Verlinde's proposal \cite{Verlinde:2016toy}, the entanglement entropy is nonzero 
only if matter is massive. 

A crucial aspect of the Verlinde proposal leading to DM identification is the presence of the de Sitter thermal bath. 
If we identify the holographic horizon \eqref{AHlate} with the entanglement horizon, it is natural to identify its 
Hawking temperature $T$ as the appropriate temperature of the de Sitter thermal bath. 
Following  \cite{Fodor:1996rf}, AH surface gravity $\kappa$ of the dynamical geometry \eqref{EFmetric}
is given by 
\begin{equation}
\kappa=\frac{d A_v}{dr}\bigg|_{r=r_{AH}} \,.
\eqlabel{defkappa}
\end{equation}
Using $\caln=2^*$ vacuum equations (4.16)-(4.22) in \cite{Buchel:2017pto} at $r=r_{AH}$ we find, 
analytically, 
\begin{equation}
\kappa=-H\,,
\eqlabel{compkappa}
\end{equation}  
independent of the mass parameters $\{m_b,m_f\}$ of the boundary gauge theory. Up to a 
sign\footnote{Presumably this is related to a negative sign in the first law in \cite{Verlinde:2016toy}.},
this is a surface gravity appropriate for the de Sitter temperature $T_{dS}=\frac{H}{2\pi}$.

In this note we reinterpreted the holographic computations of \cite{Buchel:2017pto} for the $\caln=2^*$ vacuum state in de Sitter 
 in light of Verlinde's emergent gravity proposal \cite{Verlinde:2016toy}. Using standard holographic  
gauge theory/gravity correspondence, we identified the apparent horizon of the dual geometry with Verlinde's entanglement 
horizon. We showed that the entanglement horizon has a temperature equal to de Sitter background thermal temperature.    
The Bekenstein-Hawking area of the dual AH encodes Verlinde's entanglement entropy. We emphasized that the 
entanglement entropy vanishes in the conformal limit.  If our speculation to be trusted, 
two properties of the entanglement entropy \eqref{senddef} emerge:
\nxt $s_{ent}$ is not universal --- it depends precisely on what kind of baryonic matter is entangled with the 
de Sitter gravity ---  $\calr$ depends on mass scales in the theory, see Fig.~\ref{figure1};
\nxt $s_{ent}$ is not 'linear elastic' in baryonic mass scale --- $\calr$ is a nontrivial nonlinear function of $\frac{m^2}{H^2}$.

It is interesting to further study properties of the entanglement horizon in a holographic framework. 
Presumably, turning on the mass of the boundary theory nonzero only in a finite region of the de Sitter space-time, 
it is possible to derive Verlinde's dark gravitational force.

\section*{Acknowledgments}
Research at Perimeter
Institute is supported by the Government of Canada through Industry
Canada and by the Province of Ontario through the Ministry of
Research \& Innovation. This work was further supported by
NSERC through the Discovery Grants program.

\end{document}